**Title**
- **Rules for dissipationless topotronics**
- 

**Authors**

Qing Yan, [1†] Hailong Li, [1†] Hua Jiang, [2,3]* Qing-Feng Sun, [1,4] X. C. Xie[1,2,3,4]*

**Affiliations**
1. International Center for Quantum Materials, School of Physics, Peking University, Beijing 100871, China
2. Interdisciplinary Center for Theoretical Physics and Information Sciences (ICTPIS), Fudan University, Shanghai 200433, China
3. Institute for Nanoelectronic Devices and Quantum Computing, Fudan University, Shanghai 200433, China
4. Hefei National Laboratory, Hefei 230088, China
   * jianghuaphy@fudan.edu.cn; xcxie@pku.edu.cn
   † These authors contributed equally to this work

**Abstract**

Topological systems hosting gapless boundary states have attracted huge attention as promising components for next-generation information processing, attributed to their capacity for dissipationless electronics. Nevertheless, recent theoretical and experimental inquiries have revealed the emergence of energy dissipation in precisely quantized electrical transport. Here, we present a criterion for the realization of truly no-dissipation design, characterized as $N_{in} = N_{tunl} + N_{bs}$, where $N_{in}$, $N_{tunl}$, and $N_{bs}$ represent the number of modes participating in injecting, tunneling, and backscattering processes, respectively. The key lies in matching the number of injecting, tunneling and backscattering modes, ensuring the equilibrium among all engaged modes inside the device. Among all the topological materials, we advocate for the indispensability of Chern insulators exhibiting higher Chern numbers to achieve functional devices and uphold the no-dissipation rule simultaneously. Furthermore, we design the topological current divider and collector, evading dissipation upon fulfilling the established criterion. Our work paves the path for developing the prospective topotronics.

**Teaser**

A simple rule is proposed to reach dissipationless designs in topotronics, stressing the necessity of high Chern number materials.

**MAIN TEXT**

**Introduction**

Electrical charge transport is always accompanied by energy dissipation in the form of Joule heating, which imposes a thermal bottleneck constraining the performance of traditional nanoscale integrated circuits (*1–3*). This limitation arises from the internal heat generation within these circuit elements. The emergence of novel topological systems opens up exciting avenues for optimizing thermal management beyond the current microelectronics. A myriad of topological materials host intriguing topological states, primarily characterized by the appearance of gapless boundary states that exhibit conductivity in contrast to their insulating bulk states. These topological states demonstrate remarkable and robust transport features against defects and impurities, known as topological protection (*4–8*). As one of the most distinguished topological states, Chern insulators, e.g., quantum Hall effect (QHE) and quantum anomalous Hall effect (QAHE), demonstrate a quantized Hall resistance accompanied by zero longitudinal resistance (*9–15*). Therefore, Chern insulators featuring chiral edge modes (CEM) are believed to be zero-resistive and dissipationless, which would help address the thermal bottleneck of microelectronic technologies and thereby contribute to the emerging field known as topotronics (*15–18*).

In pursuing practical device applications, the exploration of Chern insulators has progressively spread from physics to engineering, with the particular advantage of QAHE that eliminates the need for an external magnetic field. The advancements from the magnetic doped $(Bi, Sb)_2Te_3$ family (*13, 14, 19, 20*) to the $MnBi_2Te_4$ single crystal (*21–26*), and moiré superlattice (*27–29*), etc., notably enhance the quality of samples, extend the experimental observation temperature as well as reveal the higher Chern number phases. Inspired by the theoretical design of a topological current divider (*30*), recent experiments have fabricated hetero-junctions involving Chern insulators with $C = 1$ and higher Chern numbers (*31- 33*), which reported an important realization of a functional topological platform. Following recent advancements, Chern insulators offer the potential for constructing topological devices that can process information.

Although Chern insulators were expected to feature dissipationless chiral edge modes to migrate thermal issues, recent studies have revealed that heat can still generate within these topological systems, challenging these expectations (*34-37*). Even if the resistance reaches the minimum, the well-quantized electrical signature fails to eliminate dissipation inside the device (*37*). Such investigations cast doubt upon whether it is possible to reach truly no-dissipation in topotronics while maintaining functionality, scalability, and integration. Therefore, it is essential to figure out when dissipative behavior arises in topotronics and whether there are ways to avoid it.

In this work, we propose a criterion for judging whether energy dissipation occurs inside a topological device. This criterion establishes a concise algebraic relationship among the number of modes engaged in transport. We build the rule that designing a dissipationless device entails the device reaching topological protection and satisfying the proposed criterion. By examining its applicability in prior experimental findings, we discovered notable energy dissipation inside the topological devices which are typically referred as dissipationless. Fortunately, by adhering to the no-dissipation rule, we found that minor design modifications can effectively convert these devices from dissipative to truly dissipationless and the recently fabricated higher Chern number insulators play an indispensable role. Furthermore, by utilizing higher Chern number insulators as building blocks, we devise topological devices that manifest the functionality of current dividing and collecting, serving as a paradigm for dissipationless topotronics.

## Results

### Criterion for topological dissipationless devices

We present a concise overview of energy dissipation mechanisms for propagating carriers, as illustrated in Fig. 1A. Electrons are emitted from a source and injected into the device through an injecting mode. These electrons maintain equilibrium with the source. Then electrons propagate and potentially undergo scattering events due to impurities or disorder. Ultimately, two kinds of pathways emerge for their propagation: either tunneling into the drain or reverting to the source, denoted as the tunneling process or backscattering process. Notably, even though the energy of the electrons remains unchanged before and after scattering, their distribution experiences a sudden change once they are partially scattered towards either the drain or the source (*36, 37*). While the initial energy level is fully occupied, the partially occupied energy levels signify a nonequilibrium distribution (see Fig. 1A). As these non-equilibrium electrons propagate, they dissipate energy through non-elastic scattering processes, like electron-phonon interactions, resulting in heat generation within the device.

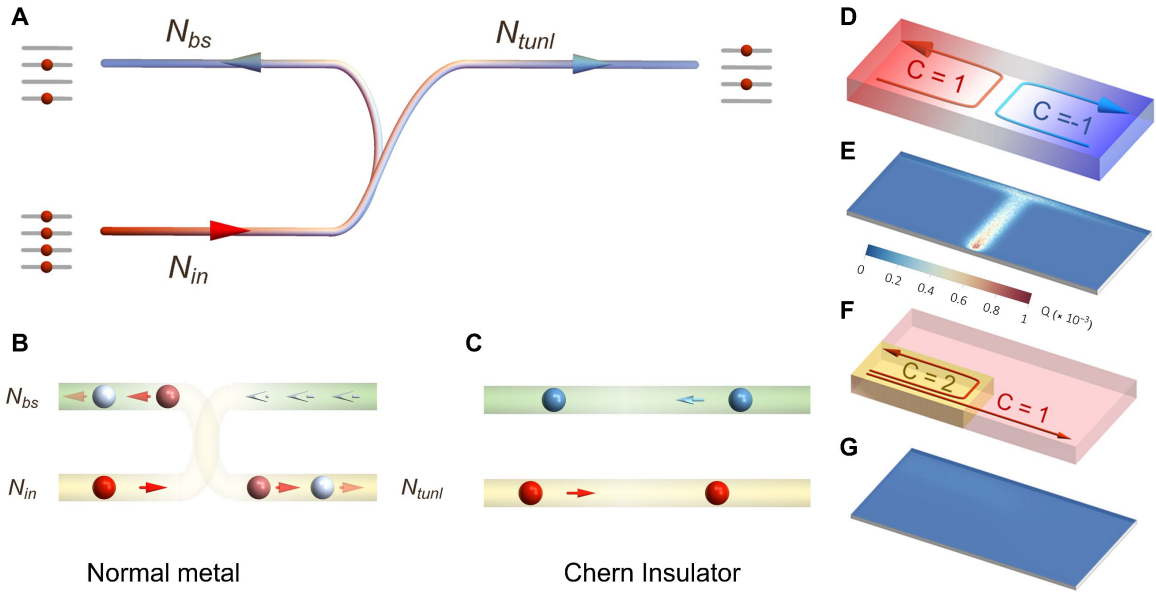

**Fig. 1. Rules and illustration of dissipation in topological devices.** (**A**) depicts the injecting, tunneling and backscattering process within a device, where the involved modes are denoted as $N_{\text{in}}$, number of injecting modes from sources in equilibrium, $N_{\text{tunl}}$, number of modes tunneling into the drain and $N_{\text{bs}}$, number of modes backscattering into the source out of equilibrium. Dissipation occurs in normal metals in (**B**) due to the broken criterion in Eq.1 with $(N_{\text{in}}, N_{\text{tunl}}, N_{\text{bs}}) = (1,1,1)$, while dissipation is avoided in the Chern insulator in (**C**) because it obeys the criterion with $(N_{\text{in}}, N_{\text{tunl}}, N_{\text{bs}}) = (1,1,0)$. For a junction of Chern insulators with $C = 1$ and $C = -1$ in (**D**), equilibrate electrons inject from the bottom left and depart with a certain probability of either backscattering to the left or tunneling to the right. The spatial distribution of heat obviously occurs in (**E**) since $(N_{\text{in}}, N_{\text{tunl}}, N_{\text{bs}}) = (1,1,1)$ violates the no-dissipation criterion. For a junction with $C = 1$ and $C = 2$ in (**F**), the criterion is satisfied with $(N_{\text{in}}, N_{\text{tunl}}, N_{\text{bs}}) = (2,1,1)$, resulting in no dissipation, as calculated in (**G**).

Driven by the need for dissipation-free devices, we introduce a definitive no-dissipation criterion that prohibits the aforementioned energy dissipation process

$$N_{\text{in}} = N_{\text{tunl}} + N_{\text{bs}} \tag{1}$$

where $N_{in}$ represents the count of modes injected from all sources, $N_{tunl}$ denotes the count of modes engaged in tunneling towards the drains, and $N_{bs}$ signifies the count of modes that retrace their path to the source. This criterion emphasizes that all the above participating modes maintain equilibrium, ensuring the absence of dissipation. Proof of the criterion is provided in **Materials and Methods**.

Here, the assertion of no dissipation is specifically confined to the region within the device. Compared to the heat generated at terminals, which can be conducted away via enlarging contact sizes to provide additional thermal cooling paths, it is more critical to address the heat generated inside the device (*2, 3*). Given the inevitable scattering centers and dissipative sources present inside actual devices, electrons consistently undergo energy dissipation during propagation, which leads to temperature increments. Such internal heat generation can notably impact the performance and reliability of electronic devices, giving rise to the bottleneck in the integration of computing elements based on normal metals or semiconductors (*2, 3*).

Equation 1 demonstrates the inevitability of dissipation in conventional electronics and elucidates the potential of topology in addressing the thermal issue. During the scattering process in normal metal, the potentially participating injecting and backscattering modes overlap spatially, i.e., $N_{bs} = N_{in}$ with $N_{tunl} \geq 1$, thereby breaking the criterion. As illustrated in Fig. 1B, the number of participating modes is $(N_{in}, N_{tunl}, N_{bs}) = (1,1,1)$, which violates the criterion outlined in Eq. 1 and brings energy dissipation. Conversely, when topology eliminates backscattering modes in Chern insulators, i.e., $N_{bs} = 0$ and $N_{tunl} = N_{in}$, the no-dissipation criterion is satisfied. In Fig. 1C, the presence of a single topologically protected channel leads to $(N_{in}, N_{tunl}, N_{bs}) = (1,1,0)$, adhering to the criterion and thus evading dissipation.

These contrasting examples seem to validate the intuitive concept of "no backscattering, no dissipation" in topological systems. However, recent experimental and theoretical studies have uncovered a series of counterexamples where heat appears in topological regimes, regardless of the presence of backscattering (*34, 36, 37*). In this context, our proposed criterion remains valid. A pertinent study in Ref. (*37*) reports a case with no additional resistance but with evident dissipation. Although the backscattering process is prohibited ($N_{bs} = 0$), the number of tunneling modes exceeds that of injecting modes, $N_{tunl} > N_{in}$. Thus, $N_{tunl} + N_{bs} > N_{in}$ leads to a scenario that there is no backscattering but still exhibits dissipation in topological systems.

Besides homogenous Chern insulators, experimental verification has validated the capacity of fabricated hetero-junctions which makes the functionalization of topological devices into reality (*31, 32, 38*). The criterion in Eq. 1 helps to judge whether the junction dissipates or not. On the one hand, the junction of $C = \pm 1$ Chern insulators can manipulate electrical current carried by CEMs which is expected to be dissipationless (*38*). However, we find that such a device exhibits obvious energy dissipation, as depicted in Fig. 1(D and E). The spatial heat distribution illustrates that even when the electron travels along the chiral edge modes, it can still undergo energy relaxation and consequent dissipation. Despite the electrical transport being topological protected, this situation breaks the no-dissipation criterion with $(N_{in}, N_{tunl}, N_{bs}) = (1,1,1)$, and thus heat generates inside the device. On the other hand, Chern junctions can be truly dissipationless once they fulfill the criterion. When a $C = 1$ regime is combined with a $C = 2$ regime in Fig. 1F, it fulfills the criterion $(N_{in}, N_{tunl}, N_{bs}) = (2,1,1)$, and there is no dissipation inside the device (see Fig. 1G).

# Criterion aiding in designing topological current divider

We have developed a criterion that provides insight into assessing dissipation in devices. It assists in categorizing existing devices and distinguishing dissipationless devices from dissipative devices. However, it remains unclear how to design a truly dissipationless topological device. To address the challenge, we propose the rule of designing dissipationless topotronics should meet two key conditions: i) hetero-junction devices satisfy the criterion of engaging modes outlined in Eq. 1, and ii) employ higher Chern number insulators as fundamental building blocks. To realize the potential functionality, it is necessary to introduce various hetero-junctions into the device. Under this circumstance, higher Chern number insulators are necessary to fulfill Eq. 1.

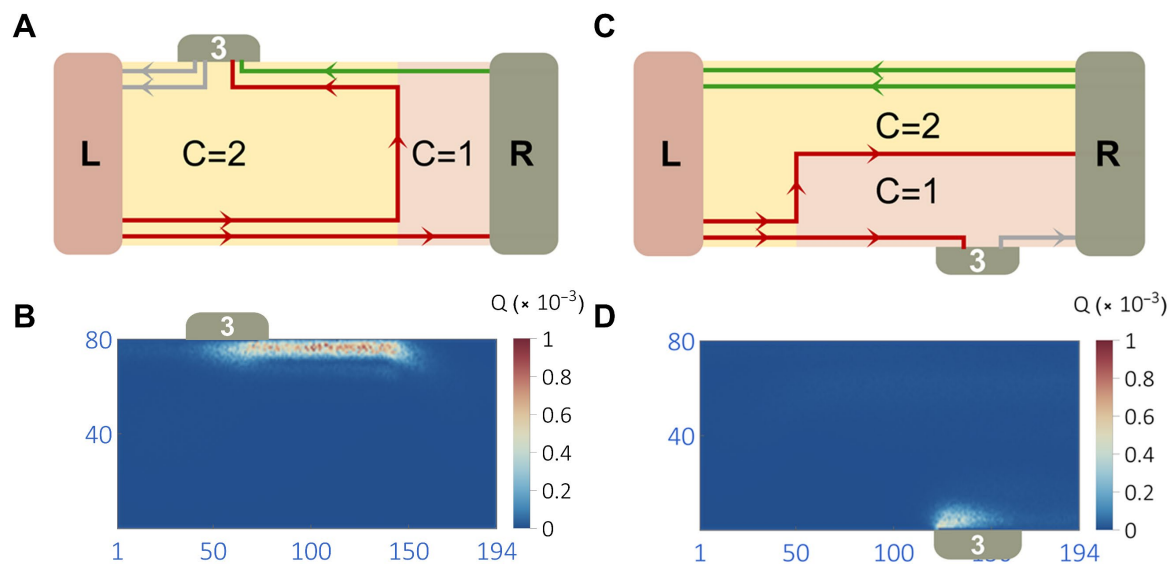

**Fig. 2. Energy dissipation occurs/disappears in topological current divider.** Two setups of topological current divider as a combination of $\mathcal{C} = 1$ and $\mathcal{C} = 2$ Chern insulators. The incident current from terminal L (as source) is equally divided into terminal R and terminal 3 (as drains) without backscattering. Lines with arrows in red (green or gray) denote the trajectory of modes injected from the source (drains). In (**A**) or (**C**), the number of participating modes $(N_{\text{in}}, N_{\text{tunl}}, N_{\text{bs}}) = (2,3,0)$ or $(2,2,0)$ violates/holds the no-dissipation criterion, which corresponds to dissipation or no dissipation in (**B**) or (**D**). The depicted energy dissipation in the two-dimensional view represents a sum over the $N_z$ layers.

The topological current divider, which has been both theoretically proposed and experimentally realized utilizing $\mathcal{C} = 1$ and $\mathcal{C} = 2$ Chern insulators, represents an important achievement in topotronics (*30-32*). However, we find that dissipation occurs even when the electrical current is distributed in a precise 1:1 ratio. The configuration in Fig. 2A replicates the device structure of Ref. (*32*) and employs an identical electrical transport measurement setup. Specifically, terminal L serves as the source, while terminals R and 3 function as drains. Note that although terminals R and 3 might appear "not active" when only considering electrical transport—since the electron injection originates solely from terminal 1—this perspective shifts when considering energy dissipation. By counting the participating modes, we find that the available tunneling modes exceed the injecting modes with $(N_{in}, N_{tunl}, N_{bs}) = (2,3,0)$ and the divider violates the no-dissipation criterion. Electrons relax and lead to evident heat generation (see numerical calculations in Fig. 2B). It confirms that energy dissipation occurs spatially inside the device before reaching terminal 3.

Using the no-dissipation criterion, we can achieve a dissipation-free topological current divider by implementing simple yet crucial modifications. The enhanced design in Fig. 2C equitably distributes current from terminal L to terminals 3 and R, which undertakes exactly the same functionality as the current divider in Fig. 2A (refer to fig. S1). Intriguingly, the tunneling modes entirely stem from the injecting modes, fulfilling the no-dissipation criterion with $(N_{in}, N_{tunl}, N_{bs}) = (2,2,0)$. Consequently, as simulated in Fig. 2D, no dissipation occurs within the device. The marginally dissipative area at terminal 3 is contact-related dissipation, which could be conducted away by high-quality contacts and does not produce harmful hot spots inside. Hence, by adhering to the dissipationless design rule, the existing device is promoted into a dissipationless version while maintaining its original function.

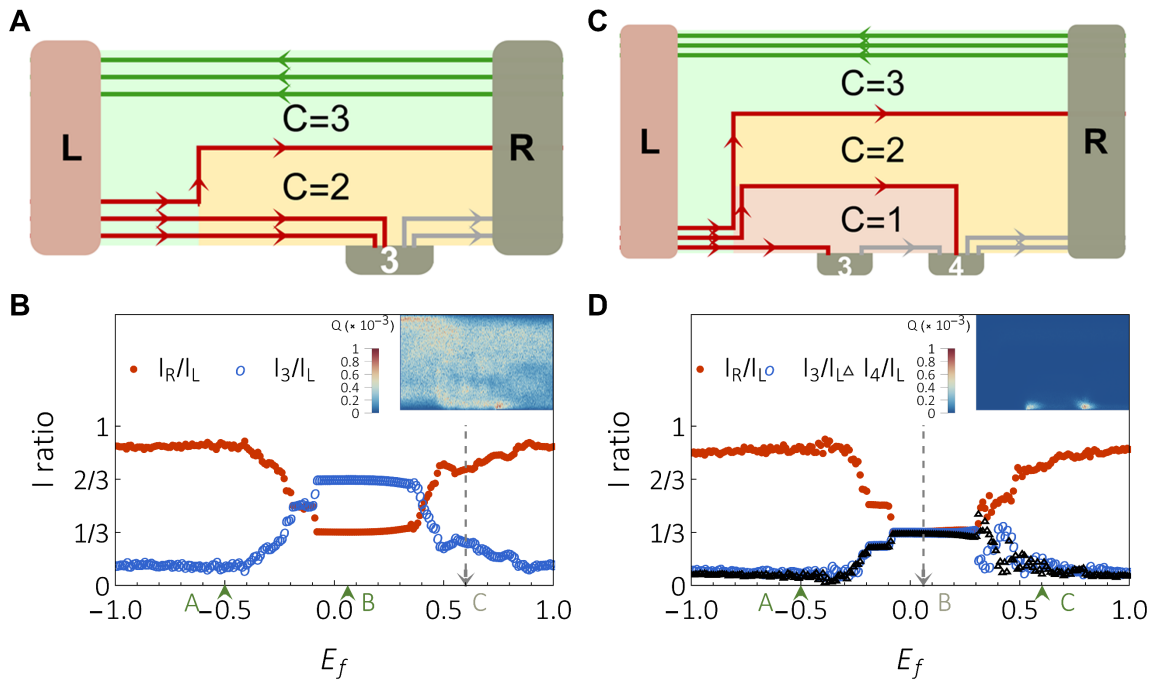

**Fig. 3. Energy dissipation occurs/disappears in topological current divider.** (**A**) or (**C**) presents two functional topological devices composed of Chern insulators with $C = 2$, $C = 3$ (and $C = 1$). Lines with arrows in red (green or gray) denote the trajectory of modes injected from the source (drains). (**B**) and (**D**) illustrate the current ratio $I_R/I_L$, $I_3/I_L$ (and $I_4/I_L$) versus the incident energy $E_f$. When $E_f$ lies in the bulk gap, the device in (**A**) or (**C**) precisely divides the incident current by the ratio 2: 1 or 1: 1: 1, respectively. In (**C**), the participating injecting modes $(N_{in}, N_{tunl}, N_{bs}) = (3,3,0)$ fulfills the no-dissipation criterion, and thus being dissipationless. The insets of (**B**) and (**D**) show the spatial heat distribution with certain $E_f$ pointed by arrowed dashed line. The depicted energy dissipation in the two-dimensional view represents a sum over the $N_z$ layers. Detailed heat distribution with certain $E_f$ marked by other points can be found in fig. S3 and S4.

Despite achieving dissipationless performance in the improved device, it is limited to a fixed 1:1 current division ratio, which constrains its application in information processing. The rule can inspire us to design current dividers with enhanced functionality. As examples, we employ $C = 3$ Chern insulators to enrich the function of current dividers.

The configuration depicted in Fig. 3A facilitates an unequal topological splitting through a $C = 3$ and $C = 2$ Chern junction. To illustrate the topological protection of electrical current, we perform current calculations relative to the chemical potential of the central region ($E_f$). These calculations reveal a scenario in which $I_R : I_3$ attains a ratio of $1 : 2$ when $E_f$ is gate-tuned within the topological gap. In this range, the participating modes adhere to the criterion $(N_{\text{in}}, N_{\text{tunl}}, N_{\text{bs}}) = (3, 2 + 1, 0)$ and there is no dissipation (see fig. S3). However, beyond this energy window, non-chiral bulk states enter the transport regime, breaking the aforementioned criterion. Consequently, energy dissipation diffuses throughout the device (refer to the inset of Fig. 3B), which provides an additional demonstration of the indispensable role of topology in dissipation-free devices.

Furthermore, the incorporation of $C = 3$ Chern insulators enables the current division into multiple terminals. In Fig. 3C, the incident current is divided into three equal portions towards three drains: $I_3 = I_4 = I_R = 1/3 I_L$, which is manifested as the current distribution in Fig. 3D. When the Fermi energy $E_f$ lies within the topological gap, the criterion is satisfied with $(N_{\text{in}}, N_{\text{tunl}}, N_{\text{bs}}) = (3, 1 + 1 + 1, 0)$. The interior of such a divider does not heat up, as evidenced by the spatial energy dissipation profile in the inset of Fig. 3D. Additional heat distributions with $E_f$ beyond the gap are illustrated in fig. S4.

Drawing inspiration from the successful applications in Fig. 2 and Fig. 3, a fascinating avenue of dissipationless topotronics emerges following the no-dissipation rule. For instance, if we employ Chern insulators with even higher Chern numbers, we can realize more complex topological dividers, such as a dissipationless divider with $n$ equal branches.

**Topological current collector**

Obeying the no-dissipation rule, we can further design more interesting and dissipationless topological devices. The current collector stands as a complementary element to the current divider, holding significance within an electrical circuit. Using higher Chern number insulators, we devise a topological current collector that includes phases $C = 1, 2, 3$. In the configuration in Fig. 4A, the incident current from three different modes is completely conducted into the right terminal. Notably, the mode count, quantified as $(N_{\text{in}}, N_{\text{tunl}}, N_{\text{bs}}) = (1 + 1 + 1, 3, 0)$, fulfills the criterion outlined in Eq. 1 and thereby eliminates all energy dissipation within the device, as depicted in Fig. 4B. Furthermore, while higher Chern number insulators can contribute to device functionality, they do not determine whether the device is dissipationless or not. For example, a device sharing identical building blocks, as depicted in Fig. 4C, serves a similar role as a current collector but it exhibits internal dissipation in Fig. 4D. The dissipation arises due to breaking the criterion with $(N_{\text{in}}, N_{\text{tunl}}, N_{\text{bs}}) = (2, 3, 0)$. This counterexample serves as a clear reminder: both aspects of the rule, satisfying the no-dissipation criterion and incorporating higher Chern number insulators as building blocks, are essential in the design of dissipationless topotronics.

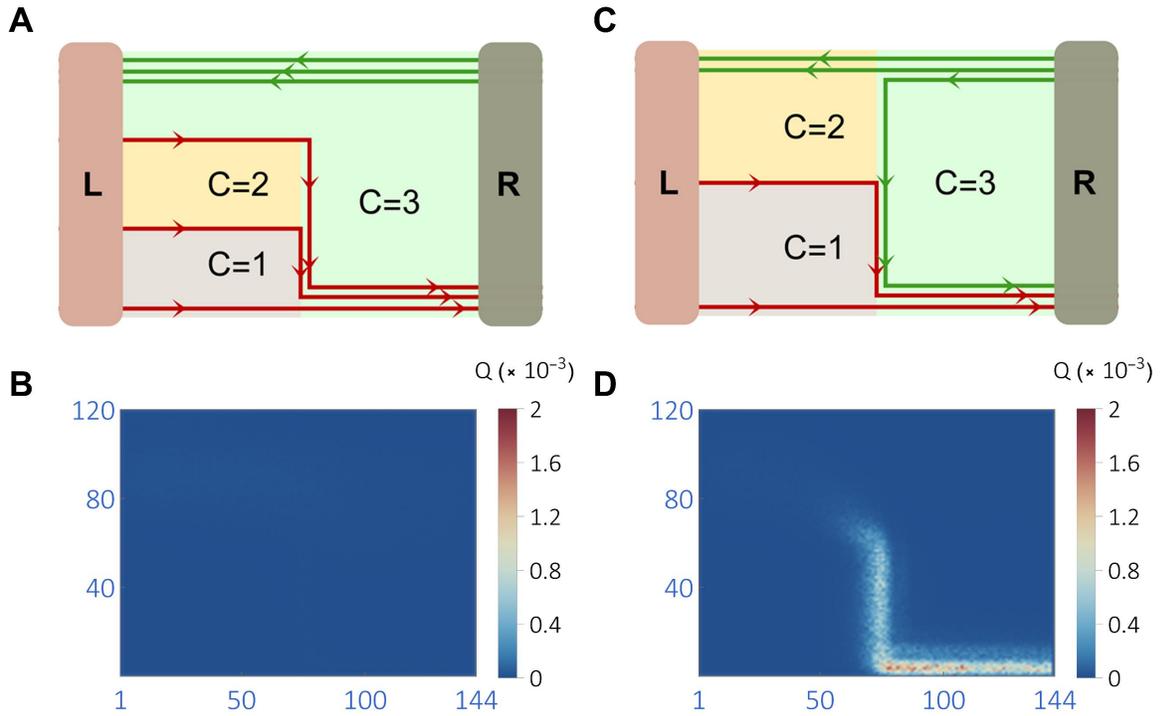

**Fig. 4. Dissipationless/dissipative topological current collectors with high Chern insulators.** (**A**) or (**C**) presents topological current collectors composed of Chern insulators with $\mathcal{C} = 3$, $\mathcal{C} = 2$ and $\mathcal{C} = 1$. Lines with arrows in red (green) denote the trajectory of modes depart from the source (drain). In (**A**) or (**C**), the number of participating modes $(N_{\text{in}}, N_{\text{tunl}}, N_{\text{bs}}) = (3,3,0)$ or $(2,3,0)$ holds/violates the no-dissipation criterion, which corresponds to no-dissipation or dissipation in (**B**) or (**D**). The depicted energy dissipation in the two-dimensional view represents a sum over the $N_z$ layers.

**Discussion**

As to the thermal challenges in microelectronics, topology provides a pathway to migrate the problem. Non-topological devices always generate heat due to scattering (as shown in Fig. 3), but topological devices offer an opportunity to reach no-dissipation. It is worth noticing that topology alone does not eliminate all heat generation. Fortunately, fulfilling the no-dissipation criterion in Eq. 1 and incorporating materials with higher Chern numbers guide the design of dissipationless topotronics.

In the context of potential platforms for exploring topotronics, recent experimental endeavors have unveiled intriguing signatures of higher Chern number insulators (*19, 20, 23, 26–28*). Possible platforms include a diverse array of materials, such as the $MnBi_2Te_4$ family, magnetic doped $(Bi, Sb)_2Te_3$ family, and moiré superlattice, etc. Furthermore, chiral semimetals represent an alternative class of topological materials that holds promise for hosting higher Chern number topological states (*39–42*). The mature techniques used in material synthesis and device fabrication pave the way for dissipationless topotronics.

Besides the aforementioned Chern insulating states, the blooming topological materials, harboring topological insulator states (*4*), valley Hall states (*43*), and high-order topological hinge states (*44*), etc., are expected to construct functional devices. At this stage, the no-dissipation criterion can be generalized to these systems. It will serve as a vital tool to assess the presence of dissipation in various topological devices, including topological-insulator-based transistors, valley filters and valves, etc.

In exploring the integration of topological systems within circuit architectures, our research reveals that functional devices comprise multiple blocks with distinct Chern numbers. As the criterion points out, achieving the desired dissipationless topotronics requires each block to operate within the topological regime. This necessitates high-quality materials and precise fabrication processes. In addition to topological systems with fermionic carriers, chiral topological superconducting systems are also pivotal for advancements in quantum computing. The transport phenomena in these systems involve not only fermionic carriers but also Cooper pairs. Further investigation should aim to include the participation of Cooper pairs when generalizing the established theory into topological superconducting circuits. Despite these limitations, our work lays a groundwork for designing integrated topological circuits towards no dissipation.

## Materials and Methods
### Effective model of high Chern insulators

We take an effective model Hamiltonian to describe Chern insulators based on a three-dimensional topological insulator model (*45–47*). Specifically, the homogeneous Hamiltonian in **k**-space is $H(\mathbf{k}) = \epsilon_0(\mathbf{k}) + \mathcal{B}(\mathbf{k})\sigma_0 \otimes \tau_z + A_2 k_x \sigma_x \otimes \tau_x + A_2 k_y \sigma_y \otimes \tau_x + A_1 k_z \sigma_z \otimes \tau_x + \mathcal{M}\sigma_z \otimes \tau_0$, where $\epsilon_0(\mathbf{k}) = C + D_1 k_z^2 + D_2 k_\perp^2$, and $\mathcal{B}(\mathbf{k}) = B_0 - B_1 k_z^2 - B_2 k_\perp^2$ with $k_\perp^2 = k_x^2 + k_y^2$. Here, $\sigma_{x/y/z}$ and $\tau_{x/y/z}$ are Pauli matrices in spin and layer degrees of freedom. Material parameters are picked as $A_1 = 2.2$ eV Å, $A_2 = 4.1$ eV Å, $B_0 = 0.28$ eV, $B_1 = 10$ eV Å$^2$, $B_2 = 56$ eV Å$^2$, $C = -0.0068$ eV, $D_1 = 1.3$ eV Å$^2$, $D_2 = 19.6$ eV Å$^2$ (*45*). $\mathcal{M}$ denotes the magnetic term which picks a constant value $M$ for any homogeneous Chern insulator region. For numerical calculation, the Hamiltonian is discretized into $N_x \times N_y \times N_z$ lattice along three dimensions.

Varying either $M$ or $N_z$ presents two pathways for controlling the Chern number within the magnetic-doped $(\mathrm{Bi, Sb})_2\mathrm{Te}_3$ family, as established in prior studies (*46, 47*), and subsequently confirmed through experimental observations (*19, 20*). Moreover, such an effective Hamiltonian can also mimic the topological properties in the $\mathrm{MnBi}_2\mathrm{Te}_4$ single crystal.

In the numerical calculation, we maintain a constant thickness along the $z$ direction and adjust the Chern number by altering the magnetic term $M$. With $N_z = 10$ and the lattice constant $a_0 = 2$, the selection of magnetic terms such as $M = \pm 0.3$ eV, $M = 0.76$ eV, and $M = 1.26$ eV, can induce transitions in the Chern insulator, resulting in phases with $\mathcal{C} = \pm 1$, $\mathcal{C} = 2$, and $\mathcal{C} = 3$.

### Transport calculation by NEGF

The numerical transport is calculated via the non-equilibrium Green's function (NEGF) combined with Landauer-Büttiker formula,

$$I_m = \frac{e^2}{h} \sum_n T_{mn} (V_m - V_n)$$
$$Q_m = \frac{e^2}{2h} \sum_n T_{mn} (V_m - V_n)^2 \tag{2}$$

where $I_m$ and $Q_m$ describe the electrical current from terminal $m$ to the central device and the energy current from the central device to terminal $m$. $V_{m/n}$ represents the voltage at terminal $m/n$. $T_{mn}$ denotes the transmission coefficient from terminal $n$ to terminal $m$ calculated by the NEGF method (*48, 49*) (also refer to Supplementary Text). Terminals $m$ take both real leads and Büttiker probes into account. Left and right leads are chosen in the semi-infinite length with the same material as the nearby central region while the additional leads, such as Lead-3 or Lead-4 in Fig. 2 and Fig. 3, are picked in a wide-band approximation (*50*). Büttiker probes mimicking dissipation sources are randomly picked in spatial with a fixed concentration $w_d$ and strength $\Gamma_d$. The energy current plotted in Figs. 2 to 4 and figs. S2 to S4 in two-dimensional view corresponds to sum over the $N_z$ layers. Details of each set-up refer to fig. S5 to S14.

**Proof of the no-dissipation criterion**

To derive the no-dissipation criterion in Eq. 1, we depict a sketch map of a general device. All the sources connected with the central region are labeled with $\alpha$ and drains with $\beta$, as shown in Fig. 5. The chemical potential of the terminal $\alpha$ is set as $\mu_h$ and that of the terminal $\beta$ is $\mu_l$. All the engaged modes in transport are grouped into three. As illustrated in Fig. 5, $N_{\text{in}}$ denotes the available modes to be populated when electrons inject from the source $\alpha$. $N_{\text{tunl}}$ denotes the available modes to be populated when electrons tunnel into the drain $\beta$. Here, $N_{\text{tunl}}$ excludes modes that directly come in and out within drains. $N_{\text{bs}}$ denotes the available modes to be populated for electrons that return back into the drain from the source.

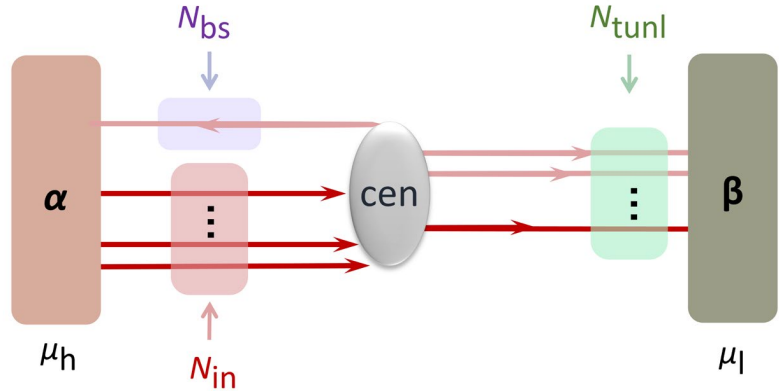

**Fig. 5. Sketch map of engaged modes.** Injecting, backscattering and tunneling modes notion are as $N_{\text{in}}$, $N_{\text{bs}}$, and $N_{\text{tunl}}$, which are boxed with color pink, green and purple, respectively. All the terminals with the same high chemical potential $\mu_h$ are regarded as sources, labeled by α, and those with the same low chemical potential $\mu_l$ are regarded as drains, labeled by β.

The derivation follows three processes that the injecting modes may participate in, which are separated by scattering and relaxation (see Fig. 6(A to G)). Injecting electrons either tunnel into sources with probability $t$ or are backscatter towards sources with probability $r$, during which process the number of particles always conserves as

$$n_{\text{in}} = n_{\text{tunl}} + n_{\text{bs}} \tag{3}$$

$$n_{\text{tunl}} = t n_{\text{in}} \tag{4}$$

$$n_{bs} = rn_{in} \qquad (5)$$

In the injection process before scattering, the number of injecting electrons per unit time is directly denoted as $n_{in} = N_{in} \int_{\mu_l}^{\mu_h} f_{in}(E)\, dE$. Here, $f_{in}(E)$ is the Fermi distribution with respect to the potential $\mu_h$, that is, $f_{in}(E) = \Theta(\mu_h - E)$ in the zero temperature (see Fig. 6A).

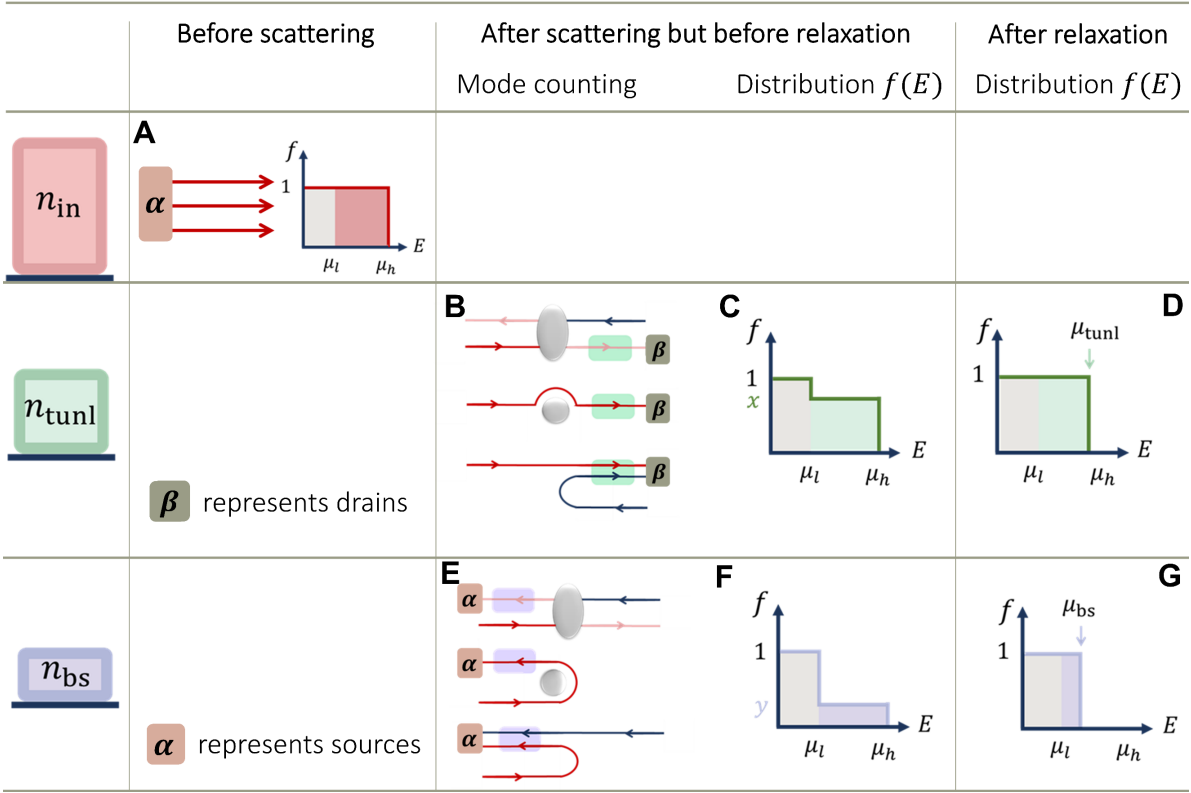

**Fig. 6. Processes of engaged modes within a device are divided into three parts.** (**A**) illustrates engaged injecting modes before scattering, demonstrating an equilibrate distribution. (**B**) depicts tunneling modes towards drains, with modes covered by green rectangles contributing to $N_{tunl}$. These modes exhibit an inequilibrate distribution prior to relaxation in (**C**) and achieve equilibrium after relaxation in (**D**). (**E**) depicts backscattering modes towards sources, with modes covered by purple rectangles contributing to $N_{bs}$. These modes exhibit an inequilibrate distribution prior to relaxation in (**F**) and achieve equilibrium after relaxation in (**G**). Shaded colorful regions in $f(E)$ between $\mu_l$ and $\mu_h$ represent the number of electrons grouped as $n_{in}$, $n_{tunl}$, and $n_{bs}$, which always fulfill the conservation law.

During the scattering process, one part of the injection electrons is scattered into the tunneling modes towards drains and their distribution function is changed into an inequilibriate form (see Fig. 6(B and C)),

$$F_{tunl}(E) = \Theta(\mu_l - E) + x\big(\Theta(\mu_h - E) - \Theta(\mu_l - E)\big) \qquad (6)$$

Correspondingly, the number of the tunneling electrons is $n_{tunl} = N_{tunl} \int_{\mu_l}^{\mu_h} F_{tunl}(E) \, dE$. Substitute it into Eq. 4 and we obtain the coefficient $x = \frac{tN_{in}}{N_{tunl}} \in [0,1]$.

Similarly, the other part of the injection electrons is scattered into the backscattering modes towards sources and their distribution function is changed into an inequilibriate form (see Fig. 6(E and F)),

$$F_{bs}(E) = \Theta(\mu_l - E) + y\big(\Theta(\mu_h - E) - \Theta(\mu_l - E)\big) \tag{7}$$

Correspondingly, the number of the tunneling electrons is $n_{bs} = N_{bs} \int_{\mu_l}^{\mu_h} F_{bs}(E) \, dE$. Substitute it into Eq. 5 and we obtain the coefficient $y = \frac{rN_{in}}{N_{bs}} \in [0,1]$.

After the relaxation process, electrons dissipate energy and reach equilibrium with the corresponding distribution function as

$$f_{tunl}(E) = \Theta(\mu_{tunl} - E) \tag{8}$$

$$f_{bs}(E) = \Theta(\mu_{bs} - E) \tag{9}$$

where $\mu_{tunl}$ and $\mu_{bs}$ denote the effective chemical potential of the tunneling modes and backscattering modes respectively. Based on the conservation condition, the re-equilibrate chemical potentials are $\mu_{tunl} = \mu_l + x(\mu_h - \mu_l)$ and $\mu_{bs} = \mu_l + y(\mu_h - \mu_l)$ (see Fig. 6(D and G)).

We can calculate the energy carried by electrons via the injecting modes, tunneling and backscattering modes as follows

$$Q_{in} = \frac{1}{h} N_{in} \int_{\mu_l}^{\mu_h} E \, f_{in}(E) \, dE \tag{10}$$

$$Q_{tunl} = \frac{1}{h} N_{tunl} \int_{\mu_l}^{\mu_h} E \, f_{tunl}(E) \, dE \tag{11}$$

$$Q_{bs} = \frac{1}{h} N_{bs} \int_{\mu_l}^{\mu_h} E \, f_{bs}(E) \, dE \tag{12}$$

The integration within Eqs. 10-12 is explicitly performed over the energy range contributing to transport phenomena, ensuring the convergence. Substitute all the distribution functions into Eqs. 10 to 12 and calculate the difference of the energy before and after the non-elastic scattering process as follows,

$$\begin{aligned}
\Delta Q &= Q_{in} - (Q_{tunl} + Q_{bs}) \\
&= \frac{1}{2h}(\mu_h - \mu_l)^2 N_{in} \left[1 - \left(t \frac{tN_{in}}{N_{tunl}} + r \frac{rN_{in}}{N_{bs}}\right)\right] \\
&= \frac{1}{2h}(\mu_h - \mu_l)^2 N_{in} \left[1 - \left(x^2 \frac{N_{tunl}}{N_{in}} + y^2 \frac{N_{bs}}{N_{in}}\right)\right] \\
&= \frac{1}{2h}(\mu_h - \mu_l)^2 N_{in} \left[\frac{N_{tunl}}{N_{in}} x(1-x) + \frac{N_{bs}}{N_{in}} y(1-y)\right] \geq 0
\end{aligned} \tag{13}$$

In the final step, we employ the conservation condition given by $x\frac{N_{tunl}}{N_{in}} + y\frac{N_{bs}}{N_{in}} = 1$ from Eq. 3, which originates from the relation $r + t = 1$. For a device exhibiting no dissipation, the condition $\Delta Q = 0$ requires

When $x = 1$ and $y = 1$,

$$t = \frac{N_{tunl}}{N_{in}}, \; r = \frac{N_{bs}}{N_{in}}, \; \Rightarrow N_{in} = N_{tunl} + N_{bs} \tag{14}$$

When $x = 1$ and $y = 0$,

$$\frac{N_{tunl}}{N_{in}} = 1, r = 0, \Rightarrow N_{in} = N_{tunl}, \; N_{bs} = 0 \tag{15}$$

When $x = 0$ and $y = 1$,

$$t = 0, \frac{N_{bs}}{N_{in}} = 1, \Rightarrow N_{in} = N_{bs}, \; N_{tunl} = 0 \tag{16}$$

Therefore, the no dissipation criterion in Eq. 1 is established.

## Acknowledgments

**Funding:**
National Key R&D Program of China grant 2019YFA0308403
National Key R&D Program of China grant 2022YFA1403700
Innovation Program for Quantum Science and Technology grant 2021ZD0302400
National Natural Science Foundation of China grant 12350401
National Natural Science Foundation of China grant 11921005
National Natural Science Foundation of China grant 12304052
National Natural Science Foundation of China grant 12374034
Strategic Priority Research Program of the Chinese Academy of Sciences National Natural Science Foundation of China grant XDB28000000
China Postdoctoral Science Foundation National Natural Science Foundation of China grant BX20220005

**Author contributions:**

Conceptualization: QY, HL, HJ, XCX
Data curation: QY
Formal analysis: QY, HL
Funding acquisition: HL, HJ, QFS, XCX
Investigation: HL, HJ
Methodology: QY, HL, HJ
Project administration: HJ, XCX
Resources: HJ, XCX
Software: QY, HL
Supervision: HJ, XCX
Validation: QY, HL, HJ, XCX
Visualization: QY, HJ
Writing - original draft: QY, HJ
Writing - review & editing: QY, HL, HJ, QFS, XCX

**Competing interests:** Authors declare that they have no competing interests.

**Data and materials availability:** All data needed to evaluate the conclusions in the paper are present in the paper, Supplementary Materials, and Zenodo: https://doi.org/10.5281/zenodo.10907366.

# Supplementary Materials

Supplementary Text

Figs. S1 to S14